\documentclass{llncs}
\usepackage{graphicx}
\usepackage{float}
\usepackage{subfigure}
\usepackage{booktabs}
\usepackage{hyperref}
\usepackage{rotating} 
\usepackage{caption}
\usepackage{multirow}
\usepackage{longtable}
\usepackage{array}

\begin{document}

\title{Security and Privacy Preserving Deep Learning}
\titlerunning{Security and Privacy Preserving Deep Learning}  
%
\author{Saichethan Miriyala Reddy \inst{1} \and Saisree Miriyala \inst{2} }
\authorrunning{Saichethan M. Reddy et al.} 
\institute{
Indian Institute Of Information Technology Bhagalpur
\and
Indian Institute of Technology Patna
\\
\email{\tt{miriyala.cse.1725@iiitbh.ac.in}}
}

\maketitle              

\begin{abstract}
Deep learning based on artificial neural networks is a very popular approach to modeling, classifying, and recognizing complex data such as videos, images, speech, and text. The unprecedented accuracy of deep learning methods has turned them into the foundation of new AI-based services on the Internet. Commercial companies that collect user data on a large scale have been the main beneficiaries of this trend since the success of deep learning techniques is directly proportional to the amount of data available for training. Massive data collection required for deep learning presents obvious privacy issues. Users’ personal, highly sensitive data such as photos and voice recordings are kept indefinitely by the companies that collect it. Users can neither delete it nor restrict the purposes for which it is used.  So, data privacy has been a very important concern for government and companies these days. It gives rise to a very interesting challenge since on the one hand, we are pushing further and further for high-quality models and accessible data, but on the other hand, we need to keep data safe from both intentional and accidental leakage. The more personal the data is it is more restricted it means some of the most important social issues cannot be addressed using machine learning because researchers do not have access to proper training data. But by learning how to machine learning that protects privacy we can make a huge difference in solving many social issues like curing disease etc. Deep neural networks are susceptible to various inference attacks as they remember information about their training data. In this chapter, we introduce differential privacy, which ensure that different kinds of statistical analysis don’t compromise privacy and federated learning, training a machine learning model on a data to which we do not have access to. 
\end{abstract}


\section{Introduction}
\label{s:intro}
Deep neural networks have shown unprecedented generalization for various learning tasks, from image and speech recognition\cite{r8} to generating realistic-looking data. This success has led to many applications and services that use deep learning algorithms on large-dimension (and potentially sensitive) user data, including user speeches, images, medical records, financial data\cite{r5}, social relationships, and location data points. Latest smart phones and devices have access to abundant amount of data which is suitable for improving i.e. training existing learning models, by which we can greatly enhance the user experience on the devices. For example, probabilistic language models can improve text input\cite{r10}, prediction and speech recognition, and similarly image models can automatically select good photos based on lightning, clarity etc.., But this data can be very personal accidental or intentional leakage can cause heavy damage to both user and organizations, so for preserving the privacy of such data conventional methods are no longer useful.

Data privacy has been a very important concern for government and companies these days. It gives rise to a very interesting challenge, since on the one hand, we are pushing further and further for high-quality models and accessible data, but on the other hand, we need to keep data safe from both intentional and accidental leakage.

When doing artificial intelligence in the real world, you'll find that most datasets are siloed within large enterprises for two reasons.

1. Enterprises have a legal risk which prevents them from wanting to share their data set outside of their organization.

2. Enterprises have a competitive advantage to hang on to large datasets collected from or about their customers.

This leads to one very challenging consequence. Scientists like us are often extremely constrained in terms of the amount of data they have access to solve their problems. These challenges are holding back possibility of active research across many problems, making it more challenging to cure disease like breast cancer, Alzheimer’s or understand complex societal trends.

The biggest tragedy of the situation is that the more personal to data and the more personal the potential uses of that data in society, the more restricted it is from scientists. This means that some of the most important and personal issues in society simply cannot be addressed with machine learning, because we do not have access to the proper training data. But by understanding how learning models work we can learn how to do machine learning or deep learning models that protects user privacy, you can make a huge difference in humanity's ability to make progress, curing disease, and truly understanding who we are through our data.
 
European Union directive that regulates the processing of personal data based on human rights law. The directive states that “[The data] controller must implement appropriate technical and organizational measures to protect personal data against accidental or unlawful destruction or accidental loss, alteration, unauthorized disclosure or access, in particular where the processing involves the transmission of data over a network, and against all other unlawful forms of processing.”

A fundamental technique for protecting private or sensitive information of individual in a data set is data anonymization. It is a procedure to erase personally identifiable information from data sets. Although it is a simple technique, it is not foolproof. For example, you can run Personally Identifiable Information (PII) such as names, social security numbers, and addresses through a data anonymization process which retains the data but keeps the source anonymous. Netflix hosted a Million-Dollar Challenge for the data science community that involved sharing movie reviews provided by 480,189 anonymous users for 17,770 movies. However few people could find a  De-anonymization Technique which used the publicly available Internet Movie Database (IMDB) as the source of data and successfully identified the Netflix records of known users which uncovered their apparent political preferences and other potentially sensitive information. Another technique, known as Differential Privacy\cite{r11}, has been receiving a lot of attention lately. In this technique, a certain amount of “noise” is added to the data without letting users know about it. The fundamental concept here is, unless anyone has access to the demising function, that party cannot retrieve the original data. Though this technique is scalable, there is some concern regarding the amount of noise that should be added to the data so that the original information is not lost to others. People have proven that differential privacy can be breached if a hacking party queries the data long enough and eventually figures out the pattern of the noise added to the data. Some researchers are working on advanced encryption techniques called Homomorphic Encryption, this is a technique where mathematical operations can be performed directly on encrypted data. When the outcome of the mathematical operation is decrypted, it is identical to the outcome of the same operation performed on un-encrypted data. This may look exciting but is yet to be stress-tested on complex and iterative mathematical operations in machine learning.

The primary cause of a privacy breach is someone’s personal data being shared with someone else. This can be avoidable by processing data in the same place that it originated. In a traditional machine learning scenario, imagine that you are developing an image recognition application. You need to transfer all the training data to a central place or in the cloud for model building. Once the model is trained, you have to send the test image back to the central server, and you get the outcome or inference on the client as shown in Figure 1(a). There are two fundamental limitations to this approach. 1. Privacy is not preserved. 2. Time lag in receiving the inference back from the cloud service. Imagine a mission-critical model like a driverless car. You cannot rely on a cloud-based service to decide whether to hit the brakes when they see a pedestrian in the car’s way. Processing capabilities in edge devices like cell phones have advanced significantly, enabling us to make inferences on the device. Here, the model is trained at a central location and deployed on the edge device. This has solved the time lag issue and has enabled cell phone applications with excellent user experience and instant inference capabilities. However, centralized model\cite{r9} training still needs all the data to be collected in a single place and is therefore vulnerable to privacy breaches. It is not possible to build a machine learning model without accessing private data.

\section{Differential Privacy}

Differential privacy\cite{r1} is a property of algorithm which limits the amount of effects and individual data
can have on the overall model output i.e., it ensures that when our neural networks are learning
from sensitive data they are only learning what they are supposed to learn without accidentally
learning from what they are not supposed to learn from data. DP typically works by adding statistical
noise either at input level or at the output level of the model or the statistical query so that you can
mask out individual user contributions but at the same time gain insights into overall population
without sacrificing privacy.

Differential privacy is about ensuring that our neural networks are learning from sensitive data, that they’re only learning what they’re supposed to learn from the data without accidentally learning what they’re not supposed to learn from the data\cite{r3}.
The general goal of differential privacy technique is to ensure that different kinds of statistical analysis don’t compromise privacy.
There are two different kinds of differential privacy techniques which refer to the two different places that you can add noise.

\subsection{Local Differential Privacy}

In this technique we add noise to each individual data point. It is adding noise directly to the database or having individuals add noise to their own data\cite{r13} before even putting it into the database. In this setting, users are most protected as they don't have to trust the database owner to use their data responsibly.
To explain the local model, let us consider a typical setup where we have a set of users with inputs at the bottom\cite{r2}, along with a data scientist at the top who would like to learn an aggregation on these inputs. In the middle we have an aggregator that receives the inputs from the users, does the actual computation, and sends the output to the data scientist
In terms of privacy, the goal is to ensure that no one should be able to learn “too much” about each individual user, and as such a bit of noise is added by each user before we send it to the aggregator.

\subsection{Global Differential Privacy }

The technique where noise is added to the output of the query on the database. This means that the database itself contains all of the private information and that it acts as an interface to the data which adds the noise necessary to protect each individual’s privacy.
To explain the global model, let us consider the same example, but here the aggregator or operator adds noise instead of the users. Doing so can still give the desired level of privacy towards the data scientist, yet now the users have to trust the aggregator

\section{Prior models}

\subsection{Centralized}

The traditional mechanisms for training or optimizing a model assumes that the entities performing those actions will have full access to those datasets which opens the door to all sorts of privacy risks. As deep learning evolves, the need for these mechanisms that enforce privacy constraints during the lifecycle of the datasets and model is becoming increasingly important.

\begin{figure}
\centering
\includegraphics[width=10cm]{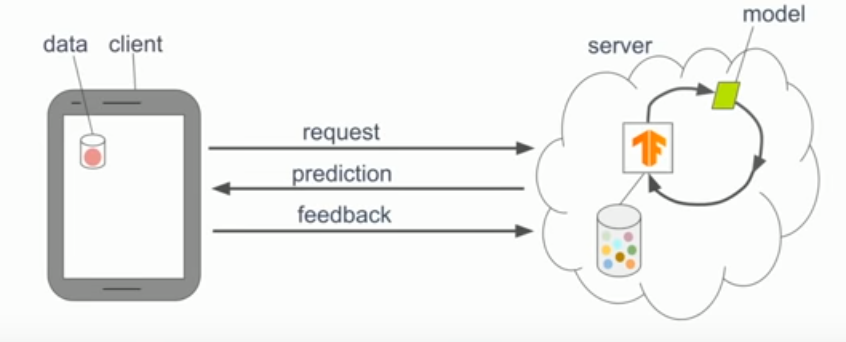}
\label{central}
\end{figure}

\subsection{Local}

In Local models, model is trained on device which preserves the data privacy but there are major drawbacks to this model such as

•	Latency

•	Battery usage

•	Computation wastage

•	Model’s will be only sub optimized

•	Models can’t be up to date

\begin{figure}
\centering
\includegraphics[width=10cm]{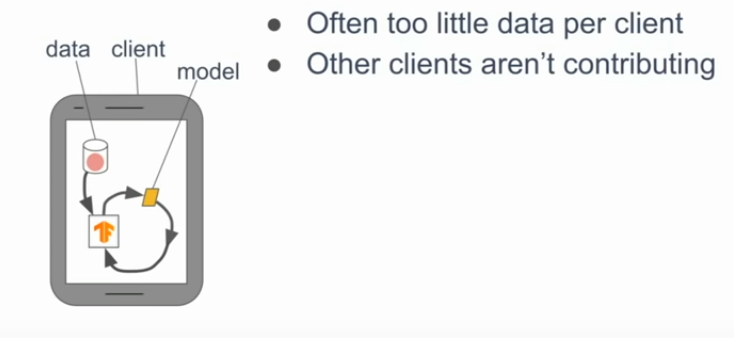}
\label{local}
\end{figure}

\subsection{Local with Proxy model}

Some of the above problems can be solved using local model with proxy data, i.e. first we train model on some proxy data then send retrained model to device by this model can me more intelligent and diverse but most of the problems will be still present.

\begin{figure}
\centering
\includegraphics[width=10cm]{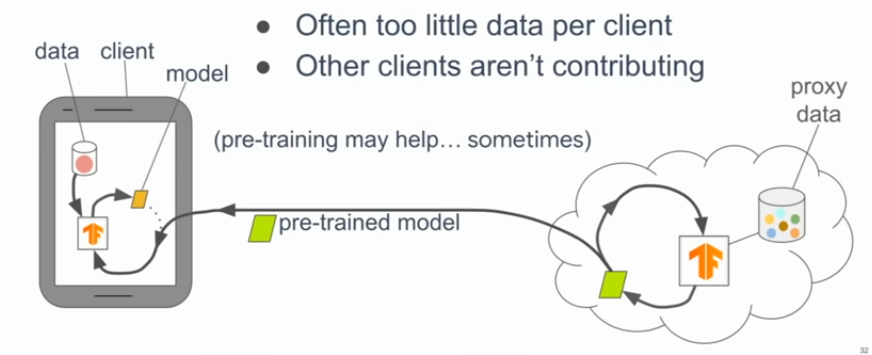}
\label{proxy}
\end{figure}

\section{Federated Learning}

\subsubsection{Definition:}
Let’s consider $N$ data holders or owners {$P1$,... $PN$ }, with respective data {$D1$,... $DN$ }. A conventional method is to collect all data together and use $D$ = $D1$ $∪$···$∪$ $DN$ to train a model MSUM (centralized approach). A federated learning system is a process in which the data owners collaboratively train a model MFED, in which process any particular data owner Pi does not expose his or her data Di to others. In addition, the accuracy of MFED, denoted as VFED, should be very close to the performance of MSUM, VSUM. Formally, let $\delta$ be a non-negative real number; if $|$$VFED$ − $VSUM$$|$ $<$  $\delta$, we say that the federated learning algorithm has $δ$-accuracy loss.

\subsubsection {Federated Learning: “Centralized model on Decentralized data”}

Federated learning\cite{r7} can be more useful in the context of mobile devices as on-device data should be used to train the central model and improve the overall model, user experience for better user experience. Previously User data is sent to server for training the model but in federated learning, the training can be brought to the device. The data is distributed across millions of devices in a highly uneven fashion which makes it even more powerful.  High level flow of federated learning on mobile devices can be summarized as follows

•	 Download copy central model 

•	training on on-device data 

•	computes the summary of changes 

•	 summaries from devices are aggregated to improve global central model 

The wealth of user interaction data on social networks, devices, including typing, gestures, video and audio capture, etc., holds the promise of enabling ever more intelligent applications. Federated Learning enables the development of such intelligent applications while simplifying the task of building privacy into infrastructure and training. Federated Learning is an approach to distributed computation in which the data is kept at the network edges and never collected centrally. Instead, minimal, focused model updates are transmitted, optionally employing additional privacy-preserving technologies such as secure multiparty computation\cite{r4} and differential privacy. Compared to traditional approaches in which data is collected and stored in a central location, FL offers increased privacy. 

\subsection{ Types of Federated Learning}

\subsubsection{Horizontal Federated Learning:}
Horizontal federated learning, systems are used when datasets share the same feature space but different space in samples. For example, two branches of bank may have very different users from their respective regions, and the intersection set of their user groups are very small. However, their business and working is very similar, so the feature spaces are the same. In horizontal federated learning system typically we assume participants are honest and security against an honest but curious server. That is, only the server can be capable of compromising data privacy of participants.

\subsubsection{Federated Transfer Learning (FTL):}
Federated transfer learning applies to the scenarios in which two datasets differ not only in samples but also in feature space. Consider two institutions: one is a bank in New York and the other is a college in southern California. Thus only a small portion of the feature space from both party’s overlaps. In this case, transfer-learning techniques can be applied to provide solutions for the entire sample and feature space. In federated transfer learning system typically involves two or more parties, so security definition for vertical federated learning can be extended here.

\section{Applications of Federated Learning}

Apart from banking, federated learning can also be effectively applied in the healthcare domain. Medical institutions relay their in-house datasets to build machine learning models. But such datasets are biased in terms of patient demographics, instruments used, or clinical specializations etc. Federated learning enables us to gain experience from a vast range of data distributed globally. Recently NVidias used Federated Deep Learning on brain tumor segmentation task and their results are quite appealing. This approach could revolutionize how the AI models are trained, with the benefits also filtering out into the broader healthcare ecosystem. Wellness apps using smartphones or wearable smart devices provides wider opportunities to leverage federated learning.

\section{GBoard - Google Keyboard, a case study}
GBoard or Google Keyboard released in 2016 is a virtual keyboard designed for iOS and android touch screen mobile devices. GBoard has over 1 billion installs as of 2019. Some of the features which make GBoard more preferred are next word prediction, auto-correction, word completion and suggesting animation, illustration based on the word. It also allows user to add emoji’s, GIFs, and stickers. 

With more than 5 billion active mobile devices, robust, fast and reliable mobile input methods become more important than ever. Features like next-word predictions provide a tool for facilitating text entry. Based on a small amount of user-generated previous text, language models (LMs)\cite{r6}\cite{r11} can predict the most probable next word or phrase. In this section, we will discuss how GBoard is using federated learning to improve query prediction or suggestions.

Before using federated learning approach, predictions were generated with a word n-gram finite state transducer. Preceding text is used for searching the next word based on the highest order n-gram state. In this problem mobile devices are referred to as clients, clients generate large volumes of personal data that can be used for training of the model. Instead of uploading user data to central servers or data warehouses for training, client’s devices train copy of global model on their local data and share model updates with the server. Weights from a large population of clients are aggregated by the server and combined to create an improved global model.
\subsubsection{Working model.}
Step A:  client devices compute Stochastic Gradient Descent updates on locally-stored on device data 

Step B:  Server aggregates the update sent by client to build a new global model 

Step C: After considerable iterations the new model is sent back to clients, and the process is repeated.

\begin{figure}
\centering
\includegraphics[width=12cm]{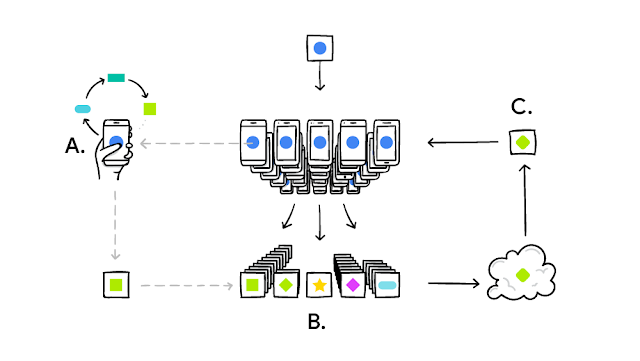}
\label{central}
\end{figure}

\section{Encrypted Deep Learning}
Encrypted Deep Learning is a technique to learn how to predict on encrypted data while the model is also encrypted. Cryptographic schemes, such as homomorphic encryption (HE), secure multiparty computation (MPC), and order preserving encryption (OPE) can ensure that analysis on data still can be performed, even without revealing information about the input on the individual level.

\subsection{Privacy-Preserving Technologies}
\subsubsection{1.	Homomorphic Encryption}
Homomorphic encryption (HE) allows arithmetic operations such as multiplication or addition on encrypted data without any need for decryption key. Formally, the encryption scheme Enc has the following equation:

\begin{equation}
    Enc(a)  Enc(b) = Enc(a ∗ b)
\end{equation}

where Enc: X → Y is a homomorphic encryption scheme with X a set of messages and
Y a set of cipher texts. a, b are messages in X, and ∗,  are linear operations defined in
X, Y, respectively.

Homomorphic encryption is a scheme which allows computation on encrypted data. In a regular environment, if one applies an algorithm on encrypted data, the decrypted output would simply be random without any meaning. With HE, decrypted result is same as if the computation was performed without encryption at all. HE is particularly valuable when outsourcing computation on the sensitive data

\subsubsection{}
For example, a hospital can outsource the data analysis on patient data to an external provider of a machine learning algorithm. The external provider won’t be able to derive any information about the patient data nor the plain result. But, they can still be able to provide a valuable analysis and insights on patients to the hospital, which has access to the plain result via its secret key.
When using HE, each time while performing an operation on the encrypted data, noise is added to the result. A scheme where the addition and multiplication operations can be applied any number of times without restriction by the noise is a certain type of HE called fully homomorphic. It is necessary to enable the computation of machine learning algorithms, but also implies a large computational overhead. This overhead makes it the slowest and least practicable scheme of the four described here. However, it is the most secure

\subsubsection{2.	Secure Multi-Party Computation }
Multi party computation (MPC) is a technique, where the computations are performed on the secret inputs from various parties. The parties gain or have access to no additional information about each other’s inputs or parameters, except from what can be learned based on output. The output is public to all parties. For example, MPC technique can be used when patient data from one hospital does not have a sufficient amount of data points to perform analysis with machine learning, but a combination of data from several hospitals has. The exchange of plain data would severely corrupt patient data privacy, however, with leveraging MPC technique it is possible to execute the required computations on encrypted data.

There are several types of MPC protocols, some based on secret sharing, some on garbled circuits etc. Simply speaking, when we apply secret sharing, a party splits their input into parts and distributes them to the other parties. These perform the computations locally, without ever seeing the actual full input values. Finally, when combining the results, the correct output is revealed. The second type of Multi party computation use logical circuits this can be imagined something like a complex binary circuit represented by a truth table. The relation between inputs and outputs is then distorted using a secret key. 

MPC is advantages over HE, such as the ability to receive input from various parties and a higher practicability due to its higher speed and less overhead. Also, MPC ensures correctness and privacy. This is a major advantage over HE which provides privacy only. However, HE is considered to have a higher level of security with less or no required communication during the computation.

\subsection{Serving Encrypted Models with Data Privacy}

Although we’ve discussed mostly data privacy, it’s also important to remember that there are at least as many reasons a stakeholder would want to maintain model privacy as well. Model privacy can be motivated by the need to secure models instilled with valuable IP, or where the models inadvertently reveal some information about its training data. Safeguarding model privacy can also help provide incentives for machine learning engineers and data scientists to develop models which they otherwise wouldn’t develop.
In situations where training with plain text doesn't work, we will need to train an encrypted model on encrypted data. This entails a non-negligible computational overhead compared to standard training, but it may be desirable when both the model and data need to be kept secret. For example, this is true in sensitive financial institutional documents or medical documents context.

\section{Conclusion}
\label{s:conc}
In this paper we tried to problems and security threats associated with existing data driven approaches. Then we introduced basics of privacy preserving techniques.



%
\bibliographystyle{splncs03}
\bibliography{splncs}
\end{document}